\documentclass[twocolumn,3p]{elsarticle}

\usepackage{color}
\usepackage{lineno,hyperref}
\usepackage{amssymb}
\usepackage{amsmath}
\usepackage{xspace}

\newcommand{\ap}{A^{\prime}}

\newcommand{\de}[1]{{\rm d}#1}

\begin{document}

\begin{frontmatter}

\title{
Diffractive photoproduction of dark photons}
\author[CVUT]{J. Cepila}
\author[CVUT]{J. G. Contreras}
\address[CVUT]{Faculty of Nuclear Sciences and Physical Engineering, Czech Technical University in Prague, Czech Republic}

\begin{abstract}
Dark photons serve as a portal to connect the dark and standard model sectors. 
Here, we propose photon-induced exclusive diffractive processes as a new tool to 
study the production of dark photons in current and future colliders. These processes join the strength of the strong force, yielding large cross sections, with the cleanliness of diffractive photoproduction, providing a clean experimental environment. Although the measurement would be extremely difficult, we identified a currently unexplored region around dark photon masses of 1 GeV/$c^2$ and coupling suppression $\varepsilon$ values around $10^{-4}$ where dark photons could be potentially found at the LHC.
\end{abstract}


\end{frontmatter}

\section{Introduction
\label{sec:intro}}

Dark matter (DM) is a compelling component of the universe used to describe, through its gravitational interactions, a large host of astrophysical observations~\cite{Bertone:2016nfn,Cirelli:2024ssz}. Nonetheless, much about DM remains unknown, including if it is made out of particles and if so, what are their properties and their interactions.
 
 One option to explain the experimental observations is to consider that the DM sector does not enter directly in contact with the Standard Model (SM) particles, but does so through a new force, see
 e.g. Ref.~\cite{Arkani-Hamed:2008hhe},  which provides a  portal joining the DM and SM sectors. One of the portals is obtained by extending  the SM with a new U(1) interaction, see e.g. Refs.~\cite{Okun:1982xi,Holdom:1985ag}, which upon mixing gives raise to a dark photon field that couples both to the SM and the DM sectors~\cite{Fabbrichesi:2020wbt}. In the minimal scenario with a dark photon $\ap$ of mass 
$m_{\ap}$, the dark photon couples to the  electromagnetic current $J^\mu$ with a suppression factor $\varepsilon$ yielding the following new terms in the Lagrangian: 
\begin{equation}
\frac{1}{2}m^2_{\ap}\ap_\mu A^{\prime\mu}+\varepsilon e \ap_\mu J^\mu,
\end{equation}
where $e$ is the electric charge.

The dark photon has been searched at accelerators using a wide variety of approaches (for a recent review see e.g. Ref.~\cite{Graham:2021ggy}); alas, without success up to now. This calls for new tools to search for it. The key idea of this proposal is to join the strength of the strong interaction, to make it more likely to produce a dark photon in the accelerators, with the cleanness of exclusive diffractive photoproduction, which  offers good experimental conditions to detect  a dark photon.

\begin{figure}
\includegraphics[width=0.48\textwidth]{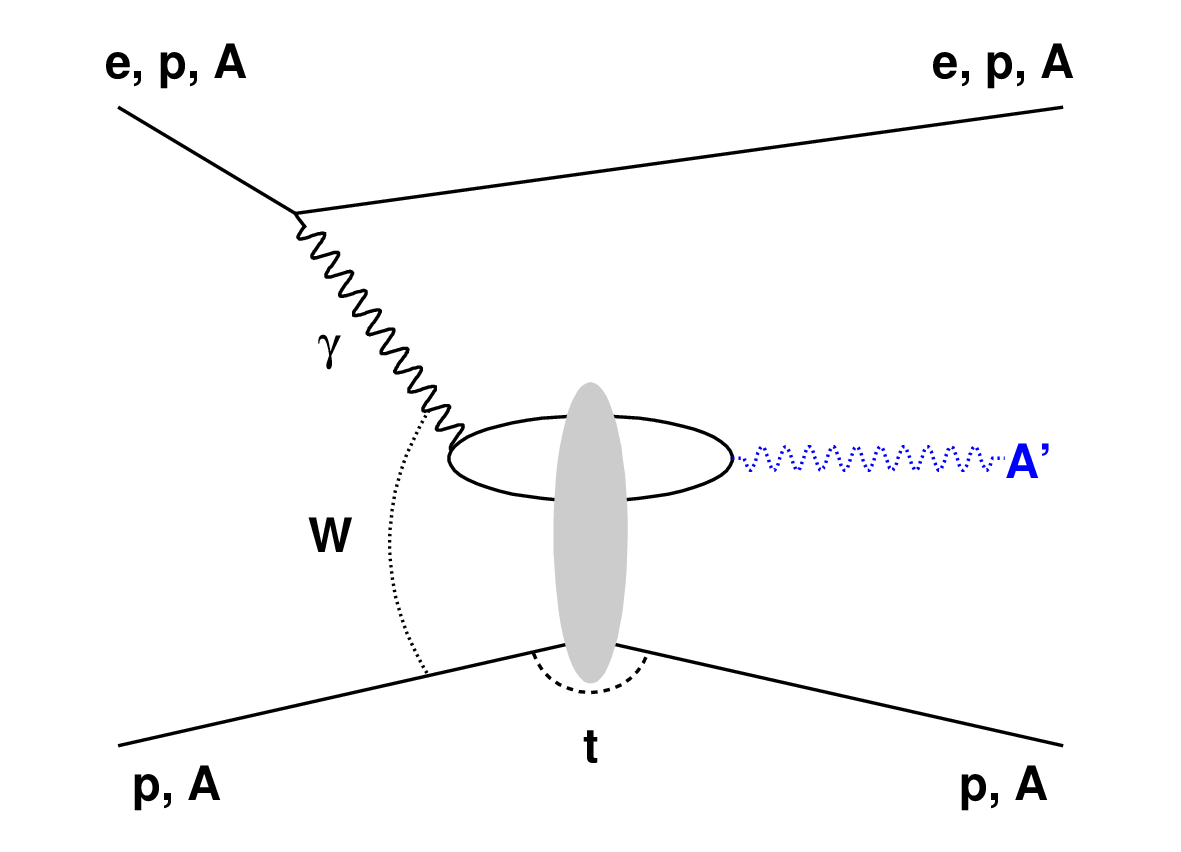}%
\caption{\label{fig:diag} Exclusive diffractive photoproduction of a dark photon. An incoming charged particle emits a quasi-real photon $\gamma$ that fluctuates into a quark-antiquark colour dipole, which after interacting with the opposite travelling particle through a colour single state (depicted as a grey oval) produces a dark photon $\ap$.}
\end{figure}

The process in question is sketched in  Fig.~\ref{fig:diag} where a quasi-real photon is emitted by an incoming beam of either electrons, protons, or nuclei. In a frame where the photon lives a long time, it may fluctuate into a long-lived quark-antiquark colour dipole which interacts diffractively with a proton or nucleus  travelling in the opposite direction through the exchange of a colour neutral, i.e. white, object---formed at leading order by two-gluons---, which is normally dubbed a pomeron; see e.g. Ref.~\cite{Kovchegov:2012mbw}.

The diffractive production of vector mesons, a very similar process to the one represented in Fig.~\ref{fig:diag} where the dark photon is replaced by a vector meson, has been extensively studied in the past at HERA in electron--proton collisions~\cite{Newman:2013ada}. Nowadays,  it is also   an important research topic at the LHC both in proton--proton and lead--lead collisions~\cite{Contreras:2015dqa,Klein:2019qfb} and it is a key component of the physics program of the EIC, currently under construction~\cite{AbdulKhalek:2021gbh}.  This process has also been widely explored from the theory and phenomenology perspective due to its importance to understand the high-energy limit of quantum chromodynamics (QCD)~\cite{Mantysaari:2020axf,Morreale:2021pnn}. The current available experience and interest, both from theorists and from experimentalists,  in the diffractive photoproduction of a vector particle, provides a fertile ground to explore its use to search for dark photons.

In this Letter, predictions are presented for  the cross section of  diffractive photoproduction of a dark photon off protons and off nuclei, Furthermore, the rates expected in current and future colliders are estimated. We identify a window in the ($m_{\ap}, \epsilon$) plane where, even though the  measurement would be extremely difficult, the dark photon could eventually be discovered at the LHC.

\section{Brief overview of the formalism
\label{sec:overview}}
The exclusive diffractive photoproduction of a vector meson---or of a real photon in deep-virtual Compton scattering (DVCS)---is explored in many works (see e.g. Ref.~ \cite{Kowalski:2006hc}).
The amplitude for the photoproduction part of the process depicted in Fig.~\ref{fig:diag} is given by
\begin{eqnarray}
&\mathcal{A}(t, Q^2, x)_{T,L} = 
i\sum_f \int \de \vec{r} \int\limits_{0}^{1} \frac{\de z}{4\pi} \int \de \vec{b} & \nonumber \\ 
& \times \left( \Psi_{\ap}^\dagger \Psi\right)^f_{T,L} e^{-i[\vec{b}- (\frac{1}{2}-z)\vec{r}]\vec{\Delta}} 2 N(\vec{r}, x),&
\label{eq:A}
\end{eqnarray}
where $f$ denotes the flavour of the quark and antiquark, Mandelstam-$t$ is the square of the momentum transferred in the photon--target interaction, $Q^2$ is the virtuality of the quasi-real photon, $T$ and $L$ denote the contributions from transverse and longitudinal photons, $\vec{r}$ is the transverse size of the dipole, $z$ the fraction of the photon energy carried by the quark, and $\vec{b}$ the impact parameter between the dipole and the target.  The exponential accounts for the Fourier transform to go from coordinate to momentum space in the transverse plane, with $\vec{\Delta}^2\equiv-t$. The dipole amplitude $N$ encodes our knowledge of QCD. In the photoproduction regime, $x=m^2_{\ap}/W^2$, where $W$ is the centre-of-mass energy of the photon--target system. The light-cone  wave functions $\Psi$  and $\Psi_{\ap}$  describe 
 the fluctuation of a photon into a quark-antiquark pair and  the process of such a pair creating a dark photon, respectively.

The overlap of the wave functions is the only one that is different with respect to the equation used for DVCS or for the diffractive photoproduction of vector mesons. For the longitudinal contribution it is given by
\begin{eqnarray}
  \left(\Psi_{\ap}^\dagger \Psi \right)^f_L
  &=& \frac{8 N_{\rm C} }{\pi}\epsilon\alpha_{\rm em}e_f^2 Q m_{\ap}z^2 (1-z)^2 \nonumber \\
 & &\times \    K_0\left(r \epsilon_\gamma \right) K_0\left(r \epsilon_{\ap}\right), 
\end{eqnarray}
while the transversal contribution is given by
\begin{eqnarray}
  \left(\Psi_{\ap}^\dagger \Psi \right)^f_T&=& \frac{2 N_{\rm C} }{\pi}\epsilon\alpha_{\rm em}e_f^2  \left\{ \right. \nonumber \\
& & 
   \left[(z^2 + (1-z)^2) \right]  \nonumber \\ 
& & \times   \epsilon_\gamma K_1\left(r \epsilon_\gamma\right)  \epsilon_{\ap} K_1\left(r \epsilon_{\ap}\right) 
    \nonumber \\
 & & + m^2_f    K_0\left(r \epsilon_\gamma \right) K_0\left(r \epsilon_{\ap}\right) \left.\right\}, 
 \end{eqnarray}
where  $N_{\rm C}$ is the number of colours, $\alpha_{\rm em}$ the electromagnetic coupling constant, $e_f$ is the electric charge of the quark in units of the electron charge, and $K_i$ are Bessel functions. Finally, $\epsilon^2_\gamma = z(1-z)Q^2+m^2_f$ and $\epsilon^2_{\ap}= z(1-z)m_{\ap}^2+m^2_f$ with $m_f$ the mass of the quark.
 The total cross section is given by the sum of the transverse and longitudinal contributions
\begin{equation}
\frac{\de \sigma_{T,L}}{\de |t|}(t, Q^2,x) =
\frac{1}{16\pi}  \left| \mathcal{A} _{T,L} \right|^2,
\end{equation}
integrated over $|t|$. 

The dipole amplitude $N(\vec{r}, x)$ is computed with the Golec-Biernat  and Wusthoff model~\cite{Golec-Biernat:1998zce} with the value of the parameters  as in Ref.~\cite{Cepila:2023dxn}. No new parameters, except $m_{\ap}$ and $\epsilon$, are needed  for the predictions of the production of dark photons presented below.

To benchmark this model, predictions for DVCS were obtained by setting   $m_{\ap}=0$ and $\varepsilon=1$. The model describes correctly the cross sections measured at HERA by the H1 Collaboration~\cite{H1:2009wnw}. The quality of the description of data is quite good, and similar to that found for other colour-dipole models as shown in Fig. 4 of Ref.~\cite{Bendova:2022xhw}.

When computing electron--proton, electron--ion, proton--proton, or ion--ion cross sections, the corresponding photon fluxes are needed. For electrons the photon flux is computed according to Ref.~\cite{Budnev:1975poe}, as customary in HERA. For the proton case the fluxes are computed
using the Dress and Zeppenfeld prescription~\cite{Drees:1988pp} with the rapidity gap survival taken from Ref.~\cite{Jones:2016icr}. For the nuclear case, the photon flux is computed as detailed in Ref.~\cite{Contreras:2016pkc}.

\section{Diffractive photoproduction of dark photons at the EIC and LHC
\label{sec:results}}

\begin{figure}[t]
\includegraphics[width=0.48\textwidth]{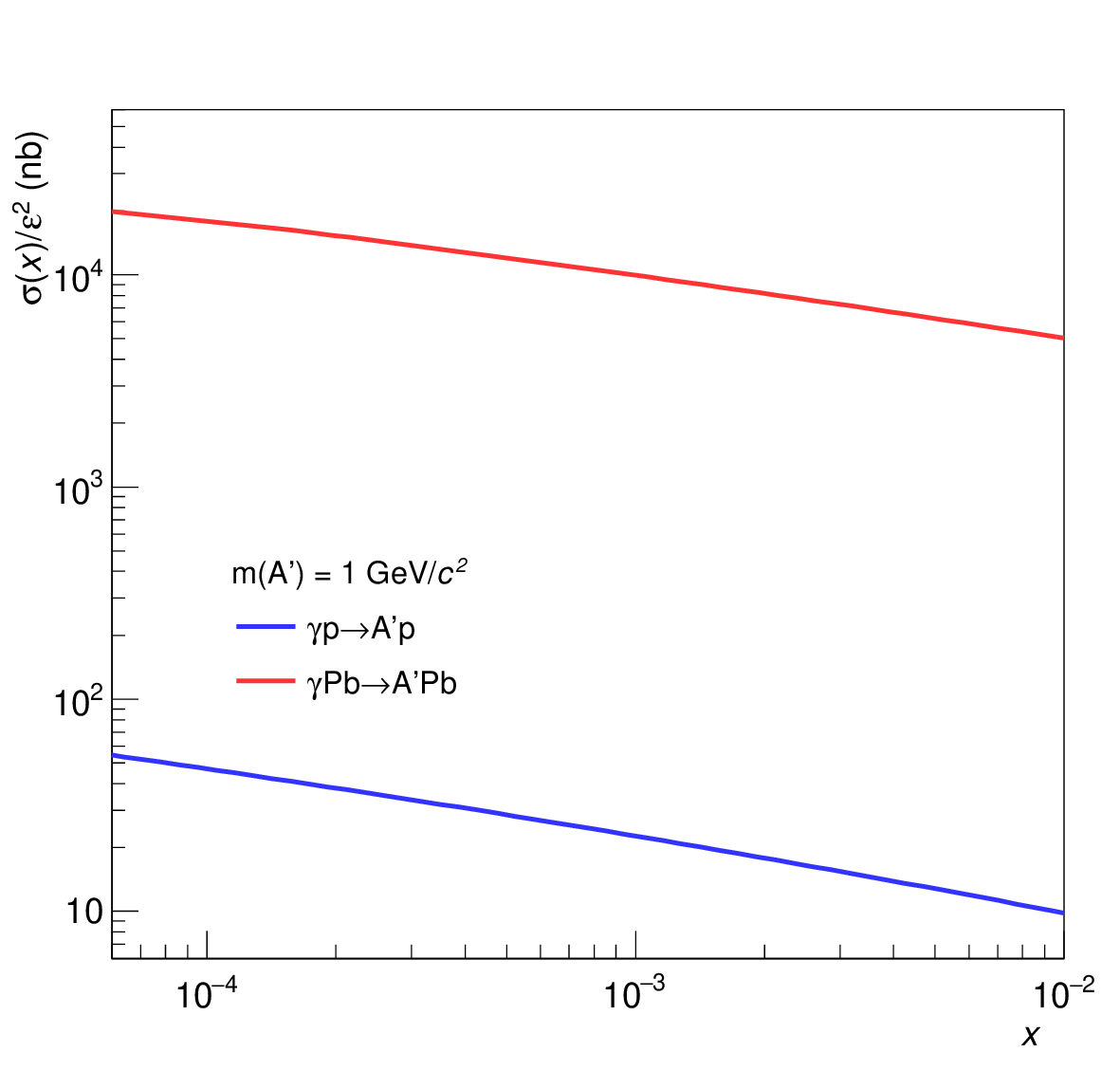}%
\caption{\label{fig:gT} Cross section, divided by $\varepsilon^2$, for the exclusive diffractive photoproduction of a dark photon of 
mass 1~GeV/$c^2$ as a function of $x$ for quasi-real photons interacting with a proton (blue line) or a Pb-ion target (red line).}
\end{figure}

\begin{figure}[t]
\includegraphics[width=0.48\textwidth]{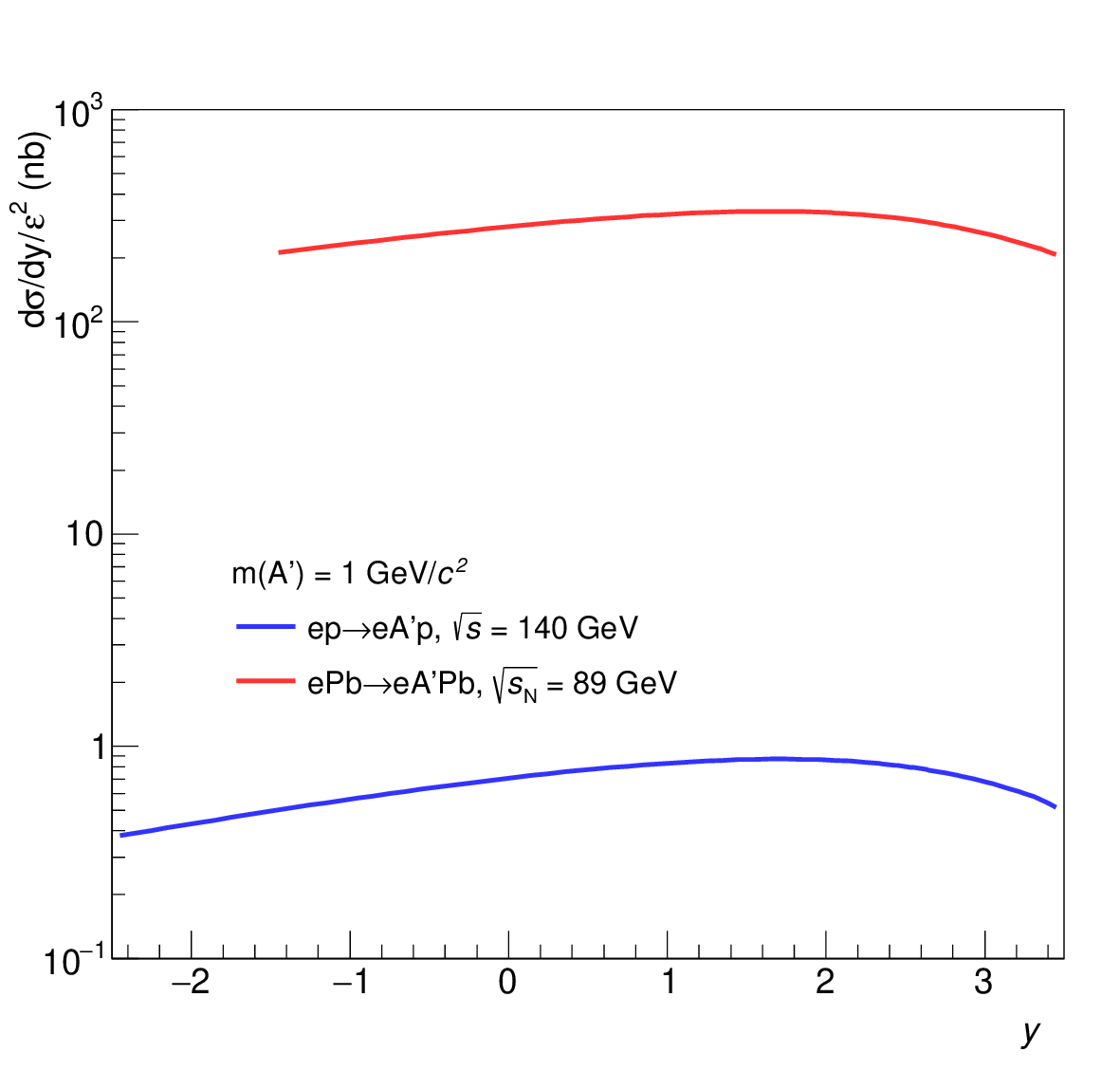}%
\caption{\label{fig:eic} Cross section per unit dark-photon rapidity ($y$), divided by $\varepsilon^2$, for the exclusive diffractive photoproduction of a dark photon of 
mass 1~GeV/$c^2$ as a function of $y$ in electron--proton and electron--lead collisions for the planned centre-of mass energies  at the EIC.}
\end{figure}

The cross section, divided by $\varepsilon^2$, for the diffractive photoproduction of a dark photon of mass 1~GeV/$c^2$ is shown in Fig.~\ref{fig:gT} as a function of $x$ for a proton and for a lead ion (Pb) target.  The cross sections are large, as expected from a process involving the strong force, and they grow as $x$ decreases. This behaviour reflects the increase of the number of gluons with the centre-of-mass energy of the interaction.  

The cross section for diffractive photoproduction off Pb is almost 3 orders of magnitude larger that than off protons. This has two reasons: first   the increase in the number of nucleons---208 for the case of Pb---, and hence gluons in the target; second, the sum of the amplitudes over the different flavours, see Eq.~(\ref{eq:A}). A similar increase on the cross section off heavy nuclei due to the contribution of different flavours is also present for the case of deeply virtual Compton scattering~\cite{Bendova:2022xhw}.

The electron--proton and electron--ion cross section, divided by $\varepsilon^2$, for the case of a Pb ion, predicted for the EIC are shown in Fig.~\ref{fig:eic}. Following Ref.~\cite{AbdulKhalek:2021gbh}, the energy of the proton (ion) beam was taken to be 275 GeV  (110 GeV per nucleon), while in both cases the electron beam has an energy of 18 GeV, producing  a centre-of-mass energy of 140 GeV and 89 GeV per nucleon, respectively. The ranges in the rapidity of the dark photon shown in the figure have been found to be reasonable for the study of ${\rm J}/\psi$ diffractive photoproduction~\cite{Lomnitz:2018juf}. The cut off of the ePb curve at lower rapidities is to restrict the predictions to values of $x$ small enough to justify the use of the dipole amplitude, which we take to be around $x=0.01$. The cross sections are relatively flat as a function the dark photon rapidity ($y$), with a slight raise at small $y$, corresponding to larger $x$, and a decrease at larger $y$ (small $x$). The decrease at large $y$ is driven by  the reduction of the photon flux for photon energies approaching the centre-of-mass energy.

\begin{figure}[t]
\includegraphics[width=0.48\textwidth]{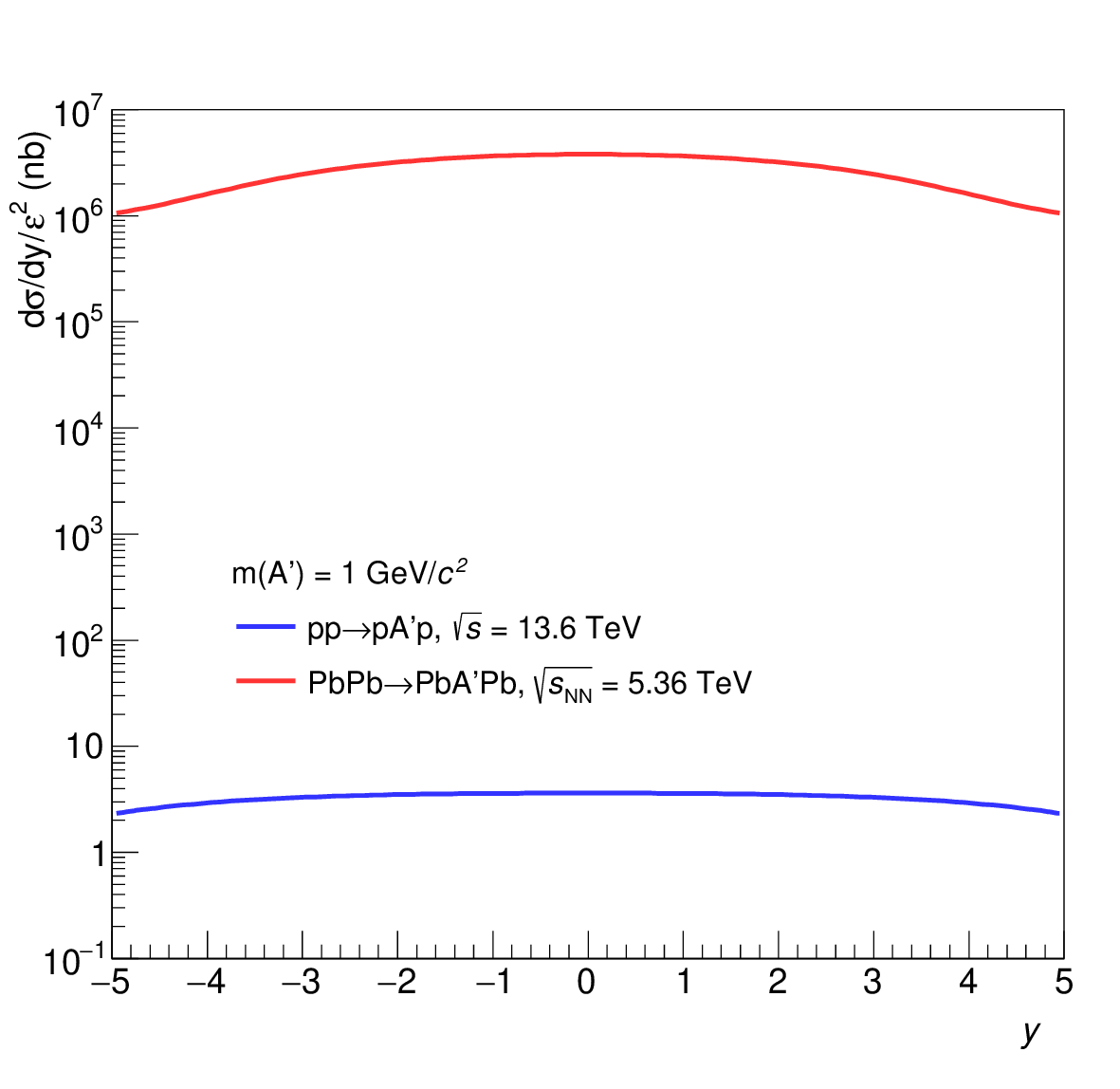}%
\caption{\label{fig:lhc} Cross section per unit dark-photon rapidity ($y$), divided by $\varepsilon^2$, for the exclusive diffractive photoproduction of a dark photon of 
mass 1~GeV/$c^2$ as a function of $y$ in proton--proton and lead--lead collisions for the current centre-of mass energies  at the LHC}
\end{figure}
\begin{table}[b!]
\caption{
\label{tab:evt} 
Expected number of events for the given integrated luminosity ($\mathcal{L}$) and range in the dark-photon rapidity ($y$) for the different collision system for EIC and LHC kinematics.}
\centering
\begin{tabular}{ccc}
\hline
System & $\mathcal{L}$ & Events$/\varepsilon^2$ \\
\hline
\multicolumn{3}{c}{EIC, $2<y<3.5$}\\
ep & 10/fb &$1.1\times 10^7$ \\
ePb & 1/fb &$4.2\times 10^8$ \\
\hline
\multicolumn{3}{c}{LHC, $2<y<4.5$}\\
pp & 3000/fb &$2.4\times 10^{10}$ \\
PbPb & 40/nb &$2.2\times 10^8$ \\
\hline
\end{tabular}
\end{table}

Figure~\ref{fig:lhc} shows predictions for the diffractive photoproduction of dark photons, divided by $\varepsilon^2$, for proton--proton
and Pb--Pb collisions at the LHC using current centre-of-mass energies. For proton--proton collisions, a rapidity gap survival factor was taken into account using the values  from Ref.~\cite{Jones:2016icr} for ${\rm J}/\psi$ photoproduction at 13 TeV.  In this case, the Pb--Pb cross section is almost six orders of magnitude larger than the proton--proton prediction. The increase with respect to that shown in Fig.~\ref{fig:gT} is due to the photon flux which depends on the square of the electric charge of the emitting particle, which for lead means an enhancement of $Z^2=(82)^2$ with respect to the flux from a proton. The cross section is again relatively flat in the rapidity of the dark photon. Due to the larger centre-of-mass energy available at the LHC, the range that can be cover by the experiments is larger than at the EIC and the full rapidity range shown in the figure fulfils that $x$ is smaller than 0.01.

\section{Expected number of dark photons at the EIC and LHC}

The expected number of produced dark photons in a specific range of rapidity for the EIC and LHC kinematics is shown in Table~\ref{tab:evt}. The quoted luminosities for the EIC are expected for one month of running, according to Ref.~\cite{AbdulKhalek:2021gbh}, so they could be scaled up according to the total integrated luminosity expected for the full lifetime of the facility. The quoted luminosity for the LHC are taken from Ref.~\cite{Aberle:2749422} and correspond to the total integrated luminosity of the HL-LHC phase, comprising Runs 4, 5, and 6 for the ATLAS or CMS experiments. The luminosity expected to be delivered for the LHCb experiment would probably smaller by around a factor of 10~\cite{CERN-LHCC-2021-012}.

The predictions have a relatively soft dependence on the mass of the dark photon.  A rough rule of thumb is that the cross section increases some 40\% from $m_{\ap}=0.5$ GeV/$c^2$ to $m_{\ap}=1$ GeV/$c^2$, with a roughly similar increase when going from $m_{\ap}=1$ GeV/$c^2$ to $m_{\ap}=1.5$ GeV/$c^2$. There is also a mild dependence in this rough 40\%, which  decreases slightly from large to small $x$. In the context of this analysis, the predictions shown for $m_{\ap}=1$ GeV/$c^2$ can be consider as representative in this mass range. In principle is possible to go lower in mass, but a dipole model is not good for such small masses and one would need to set up a vector-dominance model.

The branching ratios for the decay of a dark photon in the mass range of interest for this analysis have been studied in 
Ref.~\cite{Liu:2014cma}. Below the threshold for the decay in a muon pair, the only available channel is an electron-positron pair. 
For a dark photon mass around 500 MeV/$c^2$ the decay to $e^+e^-$, $\mu^+\mu^-$, and  $\pi^+\pi^-$ have similar branching ratios of about a third.  All three decay channels have been used at the LHC (and other facilities) to measure the diffractive photoproduction of 
$\rho^0$~\cite{ALICE:2020ugp,CMS:2019awk} and ${\rm J}/\psi$~\cite{ALICE:2023jgu,CMS:2023snh,LHCb:2022ahs}. At higher masses the 
decay into four charged pions opens up. This decay has also been measured in diffractive photoproduction at the LHC~\cite{ALICE:2024kjy}. 
The branching ratio to all this channels is above 50\% over the mass region considered here. The decay into a muon pair has already been 
used by LHCb~\cite{LHCb:2017trq,LHCb:2019vmc} and CMS~\cite{CMS:2023hwl} to set limits on dark photon production in non-diffractive 
processes. All in all, the detection of the decay products of the dark photon does not seem to be an insurmountable problem.

\begin{figure}[t]
\includegraphics[width=0.48\textwidth]{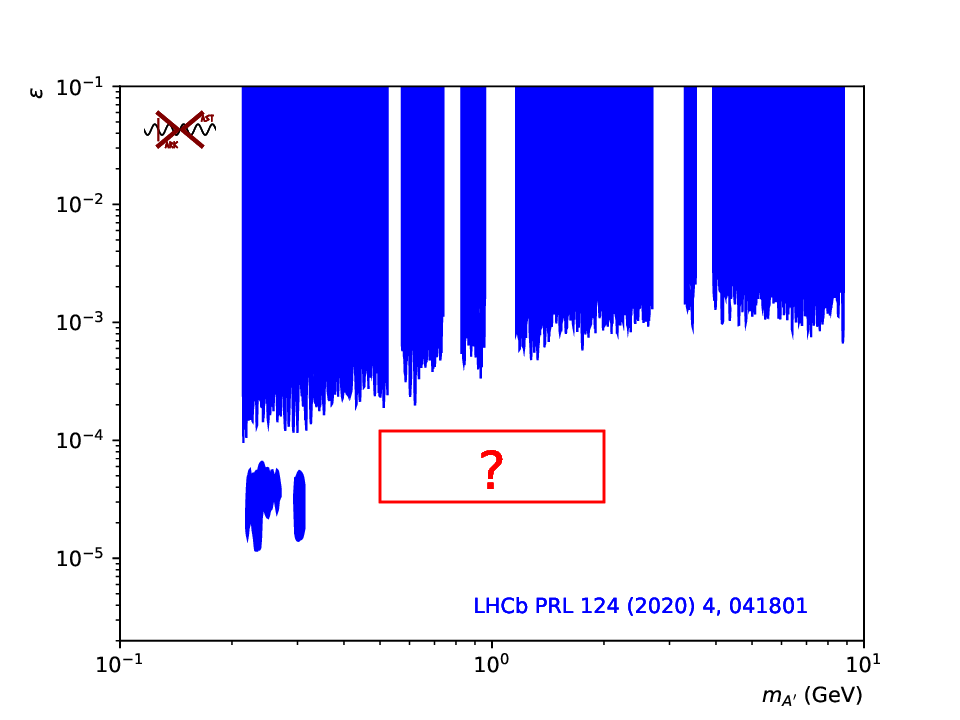}%
\caption{\label{fig:lim} Exclusion limits set by LHCb for the minimal dark photon model~\cite{LHCb:2019vmc}. The plot has been created using Darkcast~\cite{Ilten:2018crw,Baruch:2022esd}. The red box with the question mark covers a region of phase space that can be explored for the search of dark photons using diffractive photoproduction.}
\end{figure}

Nonetheless, the background could be a problem, because in addition of the production of vector mesons in this type of diffractive processes there is also production of non-resonant background, with a large cross section for the channels mentioned before. LHCb has pioneer the study of using displaced vertices~\cite{LHCb:2017trq,LHCb:2019vmc} where no irreducible background is expected.  The draw back of this approach is that it can only be used for small masses and small values of $\varepsilon$, and may need, depending on the detector, to measure at large rapidities, where there is a large boost, in order to enhance the decay length. An example of the limits set by  LHCb for this topology is shown in Fig.~\ref{fig:lim} by the blue area in the lower left side of the plot.

Given the numbers shown in Table~\ref{tab:evt}, the unexplored regions in Fig.~\ref{fig:lim}, the impressive results of the LHC collaborations in the search for dark photons, and their plans to improve their searches, see e.g.~\cite{Borsato:2021aum}, it seems possible that some new parts of the $(m_{\ap},\varepsilon)$ plane, as the one shown in red in Fig.~\ref{fig:lim}, could be explore to search for dark photons using diffractive photoproduction. It would be an extremely difficult measurement, but the LHC collaborations have shown to be capable to deliver measurements beyond the original plans for their detectors, thanks to their creativity, resourcefulness, and the excellent performance of the LHC, so such a measurement may be feasible. 

\section{Summary and outlook
\label{sec:summary}}
The cross section for the diffractive photoproduction of dark photons has been computed and used to predict production rates in electron--proton, electron--lead, proton--proton and lead--lead collisions at the EIC and the LHC. The obtained cross sections, divided by $\varepsilon^2$, are  large across a wide range of rapidity covered by the detectors of these facilities. Although the measurement would be very difficult, the expected  rates open up the possibility to search for dark photons at the LHC in a currently unexplored region of the $(m_{\ap},\varepsilon)$ plane.

\section*{Acknowledgements}
This work was partially funded by the Czech Science Foundation (GAČR), project No. 22-27262S and by the FORTE project,  CZ.02.01.01/00/22\_008/0004632, co-funded by the European Union.

\bibliography{dpdp}

\end{document}